\documentclass{aastex631}

\usepackage{amsmath,amssymb,amstext}

\newcommand{\red}{\textcolor{red}}

\shorttitle{Time-Dependent Dynamics of the Corona}
\shortauthors{Mason et al.}

\graphicspath{{./}{figures/}}

\begin{document}

\title{Time-Dependent Dynamics of the Solar Corona}

\correspondingauthor{Emily I. Mason}
\email{emason@predsci.com}

\author[0000-0002-8767-7182]{Emily~I.~Mason}

\author[0000-0001-9231-045X]{Roberto~Lionello}

\author[0000-0003-1759-4354]{Cooper~Downs}

\author[0000-0003-1662-3328]{Jon~A.~Linker}

\author[0000-0002-2633-4290]{Ronald~M.~Caplan}
\affiliation{Predictive Science Inc., 9990 Mesa Rim Rd., Ste. 170, San Diego, CA 92121, USA} 

\begin{abstract}
We present in this Letter the first global comparison between traditional line-tied steady state magnetohydrodynamic models and a new, fully time-dependent thermodynamic magnetohydrodynamic simulation of the global corona. The maps are scaled to the approximate field distributions and magnitudes around solar minimum using the Lockheed Evolving Surface-Flux Assimilation Model to incorporate flux emergence and surface flows over a full solar rotation, and include differential rotation and meridional flows. Each time step evolves the previous state of the plasma with a new magnetic field input boundary condition. We find that this method is a significant improvement over steady-state models, as it closely mimics the constant photospheric driving on the Sun. The magnetic energy levels are higher in the time-dependent model, and coronal holes evolve more along the following edge than they do in steady-state models. Coronal changes, as illustrated with forward-modeled emission maps, evolve on longer timescales with time-dependent driving. We discuss implications for active and quiet Sun scenarios, solar wind formation, and widely-used steady state assumptions like potential field source surface calculations.
\end{abstract}


\section{Introduction} \label{sec:intro}
The solar corona's structure and much of its evolution are both ultimately determined by the magnetic field at the photosphere. Phenomena such as plasma flows and magnetic reconnection are changes rooted at the solar surface; effects of these dynamics propagate into the corona, interplaying in a delicate balance and -- sometimes -- explosive loss of balance between the various forces. Remote sensing observations help researchers track and unravel many aspects of coronal dynamics, but they are often insufficient to fully understand the interplay of plasma and the magnetic field. There are many factors for this: a lack of full-Sun coverage of the evolving surface magnetic fields, the difficulties inherent in measuring coronal magnetic fields directly \citep[e.g.,][]{Dima2020}, an incomplete understanding of the solar dynamo \citep[][and references therein]{Usoskin2023}, the enormous range of scales at which critically important phenomena occur in the corona \citep{Marsch2006,Klimchuk2006,Guidoni2016}, among others. In order to delve more deeply into the mechanisms that drive coronal dynamics, we turn to models.

Magnetohydrodynamic (MHD) models of the global solar corona generally involve two elements. First a full-map of the radial magnetic field, $B_r$, in the photosphere is used set the magnetic boundary condition at the coronal base. Second, the MHD model is advanced to evolve the plasma and coronal magnetic fields until they reach a near-equilibrium state \citep[e.g.][]{Mikic1999,roussev03b,riley06_pfss,reville20}. The input magnetic map is usually static in time, constituting the so-called steady-state (SS) solution. In this case, any evolution is either due to relaxation or true coronal processes. However, when investigating large-scale dynamics over days or weeks, the time-evolution of  surface fields introduces long-period solar phenomena that are generally neglected. These include solar rotation and meridional flows, as well as global patterns of flux emergence and dispersion (and the more localized emergence and evolution of solar active regions). 

We have recently reported a time-dependently driven thermodynamic MHD simulation of the global corona covering roughtly one month of coronal evolution \red{Lionello et al. 2023}. In this Letter, we present a comparison of the time-dependent model with corresponding SS models run using the same $B_r$ boundary conditions. We find that the introduction of time-dependence has a significant effect on low-coronal dynamics, which is also apparent in the simulated emission.

In Section \ref{sec:MAS} we provide a brief overview of the model and its capabilities, as well as details of the runs. Section \ref{sec:res} presents the results of the simulations and some relevant visualizations. In the final section, we discuss what these findings mean for interpreting solar observations and for common steady-state extrapolations.

\section{MAS}\label{sec:MAS}

For our global coronal MHD calculations we employ the Magnetohydrodynamic Algorithm outside a Sphere (MAS) model. MAS solves the thermodynamic, resistive MHD equations on a nonuniform, non-adaptive spherical mesh; these equations contain coronal heating, thermal conduction parallel to field, and radiative losses. It uses a semi-implicit time-stepping algorithm, and covers the global corona and solar wind to 30 $R_{\odot}$ \citep[e.g.,][]{Lionello2013,Mikic1999,Riley2011}. A wave-turbulence-driven (WTD) approach is applied for coronal heating and solar wind acceleration to model the large-scale solar wind properties \citep{Downs2016,Lionello2014}.

The thermodynamic MHD approach allows the plasma density and temperature to be computed with sufficient accuracy to forward model EUV and soft X-ray emission and other remote sensing observables \citep[e.g.,][]{Downs2010,Downs2013,Lionello2009}. MAS has produced state-of-the-art solutions of the corona for eclipses \citep{Boe2021,Boe2022,Mikic2018}, coronal mass ejections \citep{Lionello2013,Torok2018,downs21}, and the inner heliosphere in general \citep{Riley2011,Riley2019}. Past eclipse prediction work with MAS, in particular, has shown its flexibility with generating various simulated emission quantities in forms that are easily comparable to observations. 

While quasi-steady models provide a baseline description of the solar corona, remote solar and in-situ observations over the last two decades indicate that the solar wind is always dynamically evolving \citep{Kepko2016,Rouillard2010a,Rouillard2010b}. A significant new capability has recently been added to the MAS code, detailed in \red{Lionello et al. 2023}: the ability to model the time-evolving corona in response to global photospheric magnetic flux evolution. While we lack global observations of the Sun’s surface field, flux transport models \citep{Argeetal2010,schrijver03b,upton_hathaway2014} provide us a with a sequence of global maps, as well as the underlying flows that transported the fields on the solar surface. We encourage readers to refer to \red{Lionello et al. 2023} for most of the specifics and a thorough explanation of the algorithm for this first fully time-dependent run. Here we briefly outline a few of the salient details. The simulation includes differential rotation, meridional flows, and photospheric flux emergence and decay. The evolving maps of the photospheric field are were provided by the Lockheed Evolving Surface-Flux Assimilation Model \citep{schrijver03b}, where a `synthetic' sun is simulated, with the evolving surface fields scaled and distributed such that they approximate the conditions near solar minimum. The full sequence of $B_r$ maps that were used as inputs were utilized with a cadence of one hour of solar time; there were 720 maps in total, spanning 30 days of physical evolution.



\section{Results}\label{sec:res}
\subsection{Global Evolution}

\begin{figure}
    \centering
    \includegraphics[width=0.8\linewidth,trim=200 0 0 0]{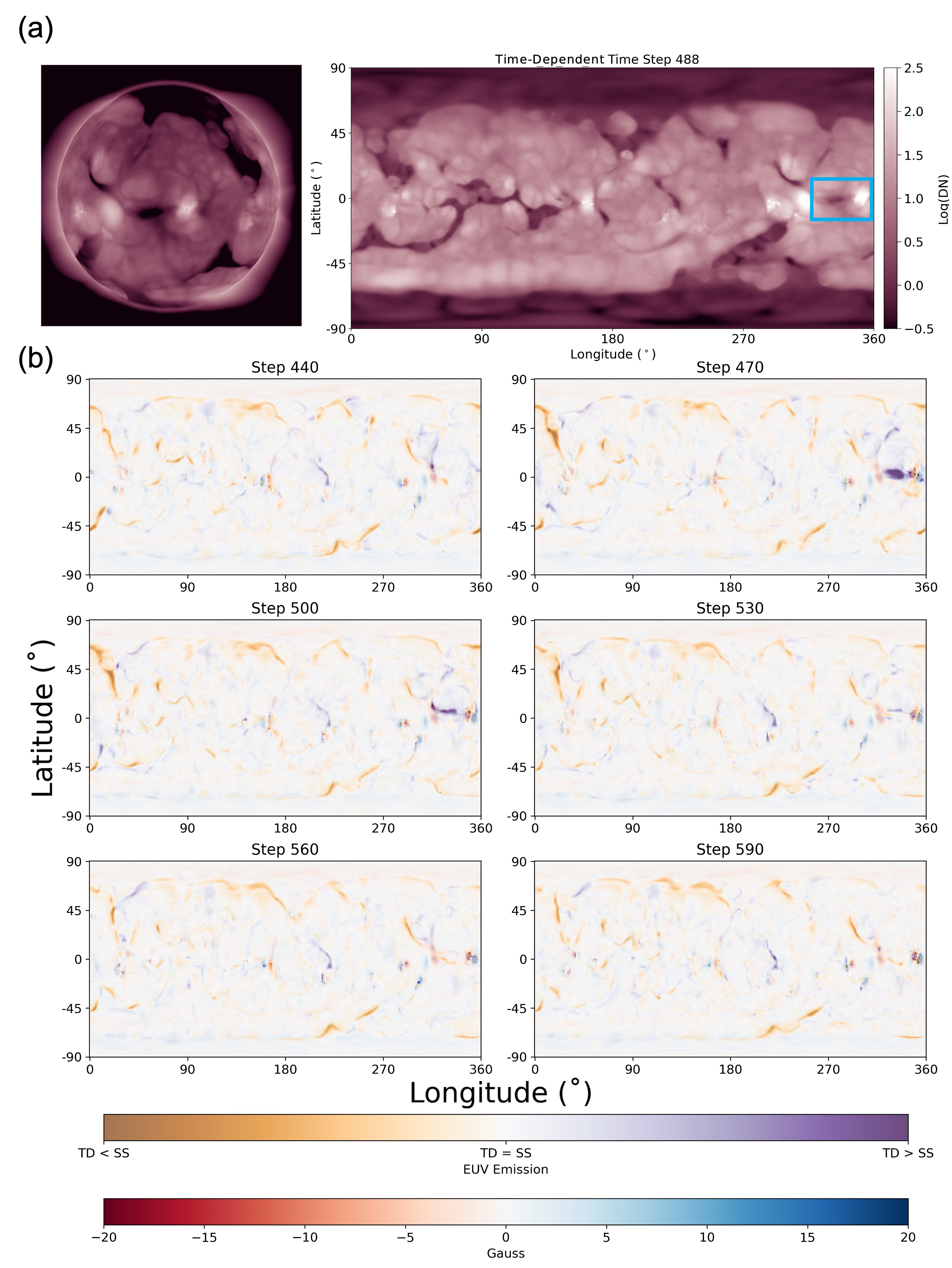}
    \caption{(a) Left: Spherical projection of the forward-modeled AIA 211 Å emission at time step 488. Right: The corresponding full-sun map of AIA 211 Å emission, integrated along the radial direction. The region highlighted in blue is the sub-region considered in the second half of this Letter. (b) Plots of $B_r$ and AIA 211 Å differenced emission; the latter color table highlights where the TD simulation emission is higher (purple) and where the SS simulations' emission is higher (orange).}
    \label{fig:euv}
\end{figure}

For this study, we compare the time-dependent (TD) simulation with 12 analogous SS simulations. The SS model runs are generated using snapshots of the full-sun surface $B_r$ at various stages in the TD model, spaced at a cadence of 15 hours between 440 and 620 hours (inclusive). Figure \ref{fig:euv}a shows an example time step from the TD run in two projections: Left, a simulated Earth view of forward-modeled 211 Å emission from the Atmospheric Imaging Assembly \citep[AIA;][]{lemen12}, which is line-of-sight (LOS) integrated along the plane of the sky. At right, we show a latitude-longitude map of the same AIA observable in Carrington coordinates, similar to a Mercator projection, where the LOS integration is instead along the radial direction. We use the AIA 193 or 211 Å channels for all of the EUV data in this Letter. The SS simulation parameters were identical to the TD run in all respects except for the evolution of the boundary conditions. The SS runs simulate just over 80 hours of coronal evolution, which is sufficient for the simulations to relax fully.

Figure \ref{fig:euv}b shows two-layer maps for a sequence of time steps showing particularly rapid evolution. The red and blue map is the photospheric magnetic flux, while the orange and purple map shows the difference in simulated EUV 211 Å emission between the TD and SS simulations of the same $B_r$ map. Orange delineates more SS emission (i.e., the TD model is darker), while purple indicates the opposite. We will discuss the sub-region with the high concentration of purple shortly; however, overall there is significantly more orange in the maps, particularly at the edges of coronal hole boundaries, consistent with the TD simulation having a higher proportion of open magnetic flux.

\begin{figure}
    \centering
    \includegraphics[width=0.55\linewidth]{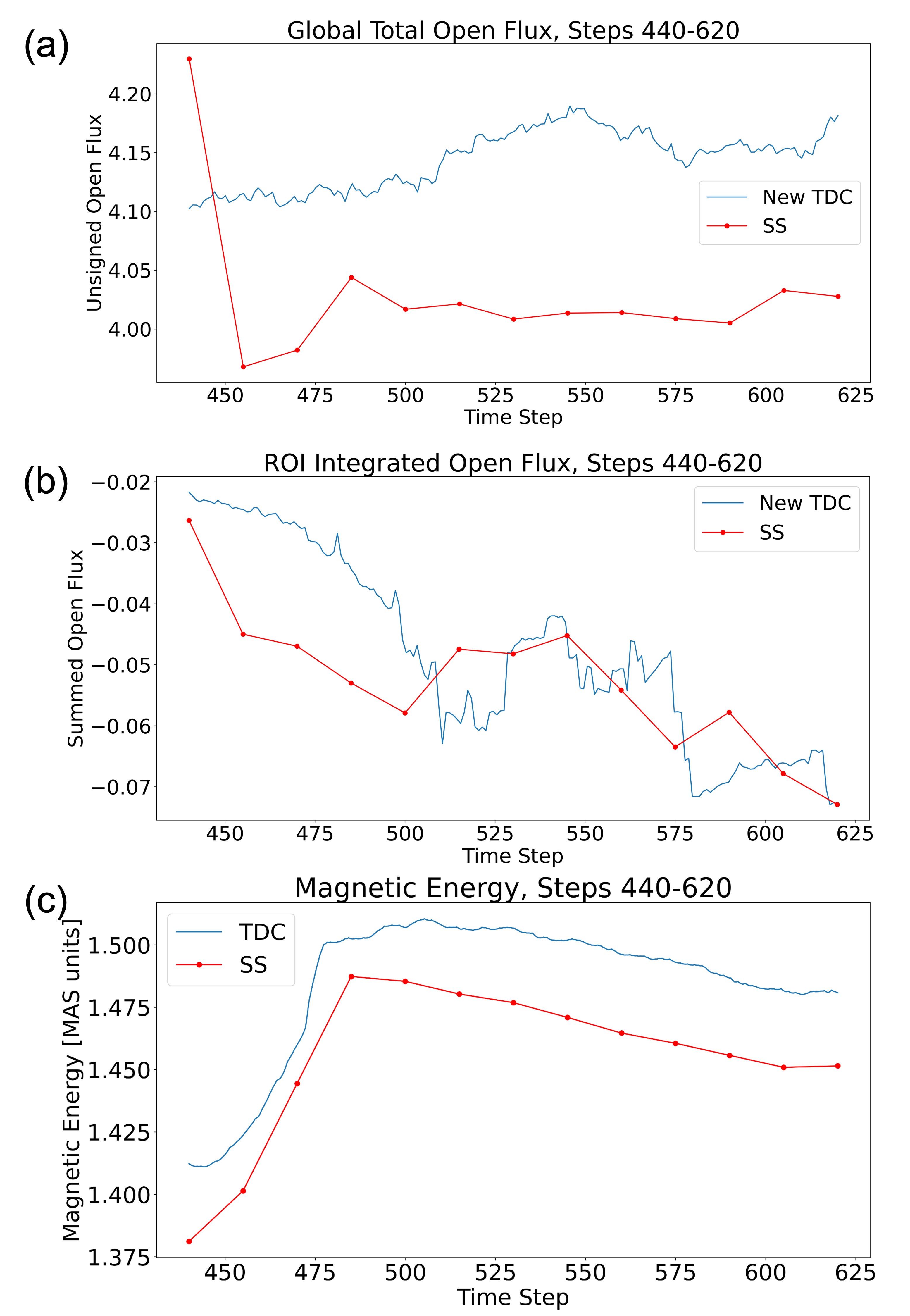}
    \caption{(a) Percent of global total open flux for SS and TD during the selected time period. Except for the first time, immediately before the emergence of the active region, the SS has uniformly less open flux than the TD simulation. (b) Comparison of net open fluxes for both cases within the region highlighted in Figure \ref{fig:euv}a. This captures the early evolution of the negative polarity coronal hole; note the much smoother and slower increase in negative flux for the time-dependent case than for the steady-state instances. (c) The  magnetic energy in both sets of simulations.}
    \label{fig:of_comps}
\end{figure}

Direct open flux calculations support this implication. As seen in Figure \ref{fig:of_comps}a, for all time steps in this interval except the first (which occurs immediately before these dynamics begin), the TD simulations consistently exhibit more open flux than the SS simulation. The additional drivers introduced via the boundary conditions for the TD simulation drive higher rates of interchange reconnection on a global scale. It also introduces more magnetic energy, as seen in Figure \ref{fig:of_comps}c. For the remainder of this Letter, we will consider this period of 180 hours between 440 and 620, which corresponds to the period of greatest flux emergence and the subsequent coronal rearrangements. In the next section, we consider these dynamics in a more specific region.

\subsection{Region of Interest Evolution}\label{ssec:roi}

\begin{figure}
    \centering
    \includegraphics[width=0.89\linewidth]{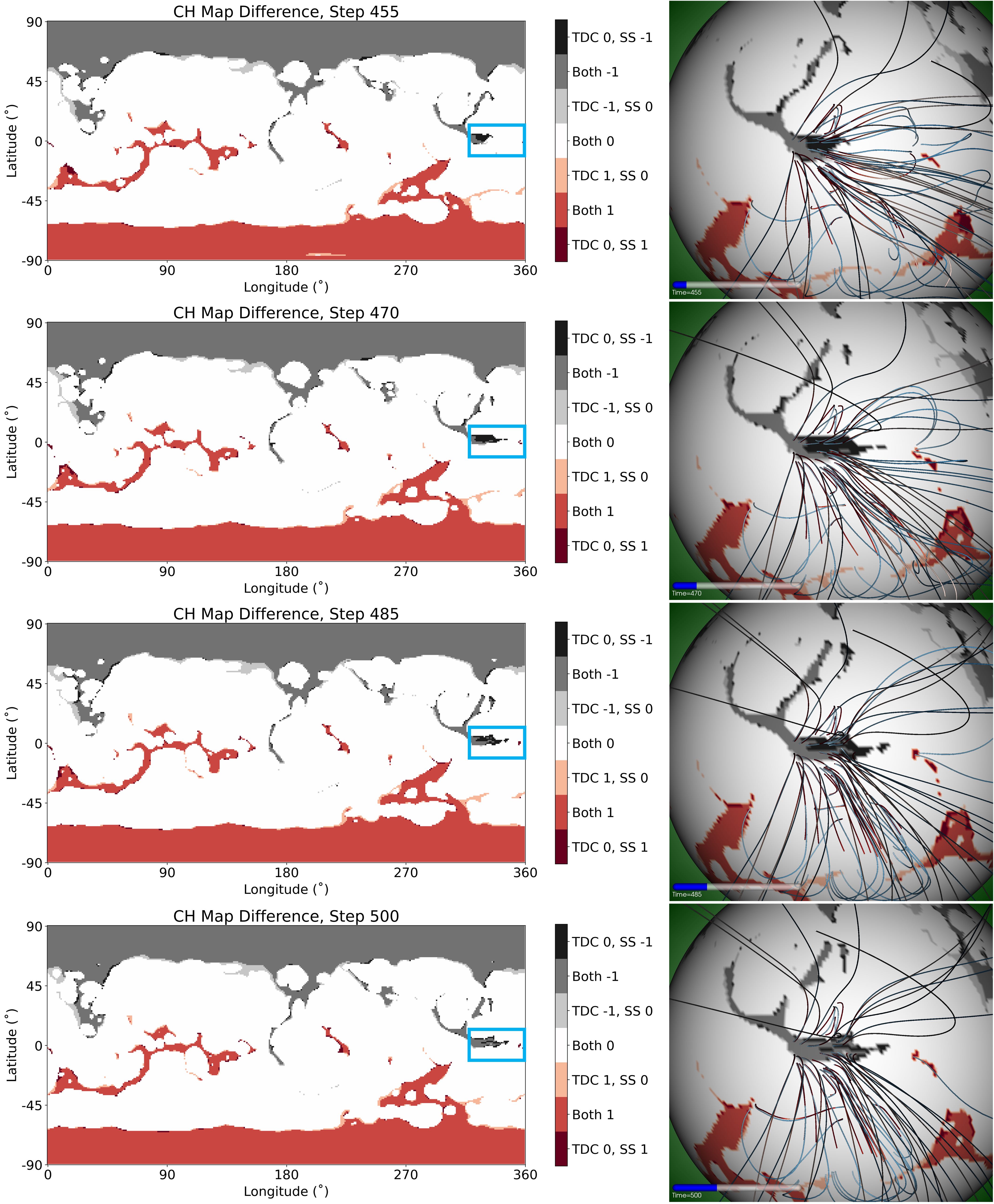}
    \caption{Left: Difference maps showing agreement and disagreement between the steady state and time-dependent runs. Black corresponds to locations where the SS is open (negative polarity CH) but the TD is closed, medium-grey is where both are open (negative polarity), light grey is where the TD is open (negative polarity) but the SS is closed, white is where both are closed, pink is where TD is open (positive polarity) but the SS is closed, red is where both are open (positive polarity), and maroon is where the TD is closed but the SS is open (positive polarity). The blue rectangle on each map shows the region integrated over for the following figure. Right: 3D visualizations of the same maps at the same times, with the SS field lines in red/orange and the TD field lines in shades of blue. The seed locations of the field lines are the same for both models, and black lines are open for both SS and TD.}
    \label{fig:chs}
\end{figure}

We studied the region highlighted in blue in the middle row of Fig. \ref{fig:chs} in greater depth (referred to as the  region of interest or ROI hereafter); during the period of hours 440 -- 620 of the full TD simulation, a large active region emerges and a coronal hole extension enlarges significantly. Figure \ref{fig:of_comps}b shows the open flux evolution during this period. We masked a rectangular area (the same area shown in blue in Figure \ref{fig:euv}a) that encompassed the expansion of the coronal hole; the SS runs exhibit more open flux faster than the TD simulation, which shows a slow but steady increase in the (negatively-oriented) open flux. Later in the evolution, the two simulations come into better agreement, with the final state values being very similar. This tendency for a significant time lag in the TD simulation is echoed in the EUV results from Figure \ref{fig:euv}. 

\begin{figure}
    \centering
    \includegraphics[width=\linewidth]{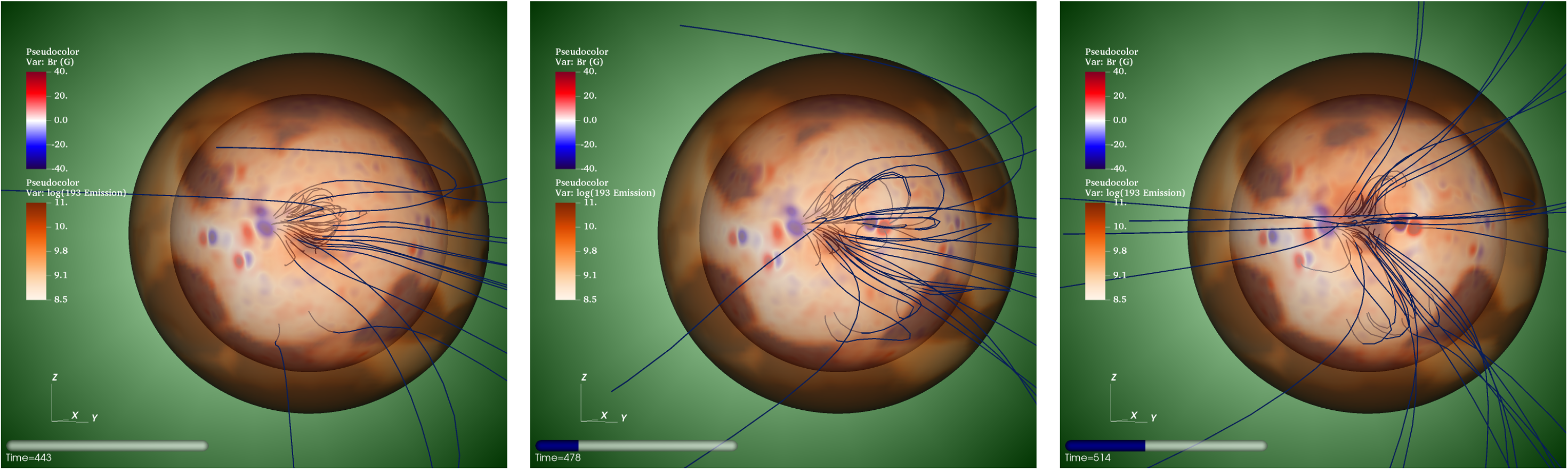}
    \caption{Visualization of $B_r$, simulated 193Å emission at 1.3 $R_\odot$, and magnetic field lines throughout the region of the emerging CH. The time stamps correspond to hours after the beginning of the simulation. Most of the field lines stretch but remain closed for over 40 hours, while some do not open until nearly 80 hours after the emergence begins. An 8-second animation of the full visualization can be found in the online version (\href{https://www.predsci.com/corona/tdc/animations/EM_2023_Fig3.mp4}{www.predsci.com/corona/tdc/animations/EM\_2023\_Fig3.mp4}).}
    \label{fig:visit}
\end{figure}

The emergence of the coronal hole in the TD simulation is presented in Figure \ref{fig:visit} (an animation showing the full time period is available in the digital version of this Letter). The visualizations show the $B_r$ at the photosphere, a spherical slice of forward-modeled 193 Å emission at 1.3 $R_{\odot}$ to show the coronal hole darkening, and field lines plotted within the ROI. The field lines stretch for roughly two days of solar time while remaining closed; the majority of this magnetic field becomes open (and correspondingly darkens in the simulated EUV) only after $\sim$50 hours post-active region emergence, while some of the field lines that eventually open remain closed for over 100 hours.

To examine the global differences between the two simulation approaches, we produced difference maps which differentiated between agreement among the TD and SS maps, and each possible type of disagreement. These difference maps are shown in Figure \ref{fig:chs}. The various shades of black and grey exhibit the open field regions for the negative polarity, while the shades of red do the same for the positive polarity. There are two important trends in the disagreement between the models: first, there is a persistent pattern of increased open field in the TD along the eastern side of most equatorial CH extensions, and increased open field in the SS along the western side. We term this pattern persistent, since it does not change greatly with time. The second trend can be seen in the ROI (highlighted with the blue rectangle). Here, the difference between the TD and SS is significant and time-dependent, during the period of active region emergence and CH expansion. The black shows where the SS model has already expanded to the full extent of the CH; this region slowly fills in with medium grey, as the TD model's field opens up and aligns with the SS field. The right column shows the same map projected onto a sphere, centered on the ROI and with field lines from both the SS and TD models to show how the fields change in 3D during this evolution.

\section{Discussion}\label{sec:disc}
In this paper, we compare a TD coronal simulation with a series of SS simulations using the same boundary conditions. The time-dependent corona results presented here include differential rotation, meriodional flows, and flux emergence and decay. The photospheric flux distribution and corresponding evolution that drive the TD model at 1 hour cadence is roughly consistent with the conditions near solar minimum. In this particular TD run, the $B_r$ evolution is matched exactly, but we do not leverage the freedom to emerge additional transverse fields and inject additional helicity along or near polarity inversion lines \citep{cheung12,Mikic2018}. Instead all evolution was due solely to the global motions and local emergence/cancellation of the radial field, making it effectively a ``minimal'' case for magnetic energy injected by driving.

To compare the differences imposed by the time-dependent model during a period of flux emergence and nearby open flux evolution, we used the same $B_r$ files for 15 evenly-spaced time steps to run steady-state simulations. These steady-state simulations were fully relaxed, but inherently lacked the magnetic ``history" of the time-dependent simulation; this was evident in systematically divergent coronal hole boundaries on the global scale, and in a time lag for localized open flux evolution on the scale of several days.

The time-dependent simulation exhibits significant temporal lags in magnetic field evolution, which affects both the local and global structure. These results show that even slow driving with relatively weak fields at the model's lower boundary leads to in several-day time differences for the open magnetic fluxes. It is particularly worth noting that the enhanced magnetic energy, while contributing to consistently  greater open flux on a global scale, does not necessarily translate to rapidly-opening flux on a local scale. This is most clearly seen in the coronal hole region focused on in Section \ref{ssec:roi}. The time lags observed here would significantly impact spacecraft connectivity calculations, modeling of co-rotating interaction region creation, and the results of solar energetic particle propagation modelling; this makes a strong argument for incorporating time-dependent driving more broadly into simulations, especially solar wind propagation models.

In the same vein, for applications such as solar wind and space weather predictions, this shows that potential-field source surface (PFSS) extrapolations which are derived from a single ``snapshot" of magnetic information may fail to represent the instantaneous structure of the corona at any moment in time. Instead, we found that the corona evinces magnetic hysteresis, where the magnetic field takes a considerable amount of time to adjust to a new configuration. Field which has been closed lengthens greatly before it opens, and this stretching can take many hours. These intermediate states, which likely constitute most of the corona for significant proportions of the solar cycle, cannot be modeled simply through an instantaneous field with no consideration of the previous state. It remains to be seen how the time lag scales with the strength of the flux which emerges; as previously noted, the emerging fields in this simulation were not particularly strong, and arguments could be made for stronger fields to generate either faster or slower evolution. The corona is likely to evolve at different rates depending upon the local topology: the presence of a nearby null point could bias the system towards faster evolution through ready interchange reconnection, while strong overlying fields connecting pre-existing active regions or the closed field of a helmet streamer may retard evolution. In order to more accurately capture the time-dependent effects in the corona discussed here, PFSS extrapolations could be adjusted to utilize a short time series of magnetic field maps and a method to interpolate the evolution between them, rather than the currently common method using a single photospheric input in time.

In addition to these hysteresis effects, there were other major differences between the SS and TD runs along the open/closed boundary for all of the associated time steps which affect traditional single-map views of open flux. Locally to the region of interest, these differences serve to illustrate the significant delay in open flux evolution in the expanding coronal hole between the SS and TD runs, as discussed above. However, the global pattern that the TD runs exhibit, with greater open flux on the eastern coronal hole boundaries and less on the western boundaries, constitute consistent and strong evidence of interchange reconnection driven by the differential rotation \citep[as popularly theorized, e.g.][and references therein]{Lionello2005}. In future simulations we plan to enhance the resolution around an open/closed boundary and study in more detail how the magnetic field and boundary evolve using tools such as the slip-back mapping method \citep{Lionello2005,Titov2009}. We will also explore the role of additional helicity injection and shear on the global coronal state. Such experiments and the use of forward modeling and observational diagnostics such as coronal hole detection and correlation dimension mapping \citep{Mason2022} will help further elucidate the role of interchange reconnection in the global corona.

Taken together, the combination of regional hysteresis and widespread constant boundary evolution shows that \emph{time dependence is important for coronal analysis on any scale and for all structures}. The quiet and active corona alike are affected by the reconnection (or suppression thereof) exhibited by the dynamics captured here. The structure of coronal holes affect the shape and evolution of the helmet streamers by which they are connected. This, in turn, modulates the processing of magnetic field throughout the small-scale quiet Sun in the low corona to the large-scale closed regions above \citep{Liu2021,Morosan2020,Schlenker2021,Scott2021}. While the total open flux for the TD run was only greater by a few percent, that few percent is concentrated in a few narrow areas along open/closed boundaries, heightening its influence there. Furthermore, as seen in the region of interest highlighted here, short-term evolution like the emergence of an active region can distort local fields rapidly, while the end-result structural changes (i.e., the expansion of the coronal hole) don't occur until days later. All of these aspects contribute to the background state of the global corona at any given time. This has major implications for impulsive events like flares and prominence eruptions, despite the fact that such events' timescales are too short to be directly affected by global details like differential rotation. Ultimately, the actual corona is evolving continuously in response to photospheric evolution: modeling which aims to explore magnetic field structure and evolution will benefit from being time-dependent as well.

\section{Acknowledgments}

This work was supported by the NASA Heliophysics Living With a Star Science and Strategic Capabilities programs (grant numbers 80NSSC20K0192, 80NSSC22K1021, and 80NSSC22K0893), the NASA Heliophysics System Observatory Connect program (grant number 80NSSC20K1285), and the NSF PREEVENTS program (grant ICER1854790). Computing resources supporting this work were provided by the NASA High-End Computing (HEC) Program through the NASA Advanced Supercomputing (NAS) Division at Ames Research Center and by the Expanse supercomputer at the San Diego Supercomputing Center through the NSF ACCESS and XSEDE programs.

\bibliography{TDCvsSS} 
\bibliographystyle{aasjournal}



\end{document}